\documentclass[aps,prd,showkeys,showpacs,twocolumn,superscriptaddress]{revtex4}
\usepackage{amsfonts}
\usepackage{amsmath}
\usepackage{latexsym}
\usepackage{amssymb}   

\newcommand{\ov}[1]{\overline{#1}}
\newcommand{\lra}{\longrightarrow}
\newcommand{\ul}{\underline}
\newcommand{\ket}[1]{| #1 \rangle}
\newcommand{\bra}[1]{\langle #1 |}

\newcommand{\trace}{\textrm{Tr}}

\newtheorem{Def}{Definition}
\newtheorem{Protocol}{Protocol}

\newtheorem{Cl}{Claim}
\newtheorem{Lem}{Lemma}
\newtheorem{Proposition}{Proposition}
\newtheorem{Scenario}{Scenario}
\newtheorem{Conj}{Conjecture}
\newtheorem{Cor}{Corollary}

\begin{document}

\title{\begin{center}
Blind quantum computation
\end{center}}

\author{Pablo Arrighi}
\email{pablo.arrighi@imag.fr} \affiliation{Laboratoire 
Leibniz, Institut d'Informatique et de Math\'ematiques 
Appliqu\'ees de Grenoble (IMAG), CNRS UMR 5522, 46 Avenue 
F\'elix Viallet, 38031 Grenoble Cedex, France.}

\author{Louis Salvail} \email{salvail@brics.dk} 
\affiliation{BRICS, Department of Computer Science, 
University of Aarhus,\\Building 540, Ny Munkegade, Aarhus 
C-8000, Denmark. }

\begin{abstract}
We investigate the possibility of \emph{having someone 
carry out the work of executing a function for you, but 
without letting him learn anything about your input}. Say 
Alice wants Bob to compute some known function $f$ upon her 
input $x$, but wants to prevent Bob from learning anything 
about $x$. The situation arises for instance if client 
Alice has limited computational resources in comparison 
with mistrusted server Bob, or if $x$ is an inherently 
mobile piece of data. Could there be a protocol whereby Bob 
is forced to compute $f(x)$ \emph{blindly}, i.e. without 
observing $x$? We provide such a blind computation protocol 
for the class of functions which admit an efficient 
procedure to generate random input-output pairs, e.g. 
factorization. The cheat-sensitive security achieved relies 
only upon quantum theory being true. The security analysis 
carried out assumes the eavesdropper performs individual 
attacks.
\end{abstract} 



\maketitle
\section{Introduction}
In the traditional secure two-party computation 
scenario\cite{Yao,Abadi2} Alice has secret input $x$, Bob 
has secret input $y$, and both of them wish to compute 
$f(x,y)$. The function $f$ is of course well-known to the 
two parties; the usual example is that of two millionaires 
who wish to compare their wealth without disclosing how 
much they own\cite{Yao}. Most protocols for secure 
two-party computation are symmetric with respect to the 
computing power each party should carry out during the 
execution. In these scenarios, if Alice knew Bob's input 
$y$ she could compute $f(x,y)$ on her own without having to 
invest more computing power. Entering a secure two-party 
computation together with Bob will in general not help in 
diminishing Alice's computing power needed to evaluate $f$, 
and this is simply not the aim pursued. In fact, all 
implementations known by the authors of this paper require 
both Alice and Bob to invest more computing power than what 
is needed for the mere evaluation of $f$. For instance in 
\cite{Abadi2} each gate performed by a party requires the 
other to perform the same gate, together with some extra 
encryption.\\

Unlike secure two-party computation, blind computation is 
fundamentally asymmetric. Alice is the only party with a 
secret input $x$, Bob is the only one able to compute $f$. 
Alice wants Bob to compute $f(x)$ without him learning {\em 
too much} about $x$. Thus an obvious motivation for Alice 
to enter a blind quantum computation together with Bob is 
to unload the computational task of computing $f$ without 
having to compromise the privacy of her input. One could 
easily imagine this occurring in a Grid architecture, or in 
any client-server relation with a mistrusted server 
retaining the computational power. To make things more 
precise, suppose there were only a handful of fully 
operational large-scale quantum computers in the world, and 
some hungry academic decided to make use of her timeshare 
as scientist to crack some Swiss bank's $RSA$ private key 
$x$. The hungry academic (Alice) will surely want to keep 
$x$ secret from the authorities handling the quantum 
computer (Bob), so that she does not get suspected when 
subsequent international money transfers come to top up her 
meager income. But there may be other reasons to enter a 
blind computation protocol than mere computational power 
asymmetry. For instance Bob may possess some trapdoor 
information about the otherwise well-known function $f$. Or 
perhaps $x$ may represent some mobile agent's code which 
ought to be protected against the malicious host upon which 
it runs. Others may see blind quantum computation as a 
somewhat philosophical issue: Is it possible to carry out 
some work for someone whilst being prevented from knowing 
what the work consists in?

In the classical setting, blind computation has first been 
studied by Feigenbaum \cite{Feigenbaum}. It was shown that 
for some functions $f$, an instance $x$ can be encrypted by 
$z=E_k(x)$ in such a way that Alice can recover $f(x)$ 
efficiently from $k$ and $f(z)$. The construction cannot be 
extended easily to general classes of functions. In 
particular, blind computation of the discrete logarithm 
function (DLF) was shown possible but no blind computation 
of the $RSA$ factoring function (FACF) is known.
%Moreover Abadi, Feigenbaum, and Kilian \cite{Abadi1} have shown
%that no NP-hard problem can be computed blindly unless the
%polynomial-time hierarchy
%collapses at the third level,
%and this seems to remain true when the privacy of Alice's input is only
%partial.
The infinite complexity hierarchy $P\subseteq 
NP=NP^{\emptyset}\subseteq NP^{NP} \subseteq \ldots 
\subseteq NP^{NP^{{.^{.}}^{NP}}}\subseteq \ldots$ (where 
$NP^{C}$ stands for the class of language recognizable in 
non-deterministic polynomial time provided access to an 
oracle for problems of class $C$) is called the {\em 
polynomial-time hierarchy}. It is widely believed that 
every level in the polynomial-time hierarchy is strictly 
contained in the next one. However, proving or disproving 
this statement would be a major breakthrough in complexity 
theory. Abadi, Feigenbaum, and Kilian\cite{Abadi1} have 
shown that no NP-hard problem can be computed blindly 
unless the polynomial-time hierarchy collapses at the third 
level. We conclude that it is very unlikely that any 
NP-hard problem can be computed blindly in the classical 
setting.

Even when computational assumptions are invoked 
\cite{Sander}, none of the currently known classical blind 
computation protocols applies to general classes of 
functions. Rather they take advantage of specific algebraic 
properties of particular functions. These constructions 
rely upon encryptions that are, in some sense, homomorphic 
with respect to function $f$. Clearly, very natural 
candidates for $f$ are not known to have this property like 
for FACF. It is not surprising that such stringent 
requirements do not necessarily hold when Bob is running a 
quantum computer.

%Whilst on the one hand secure multi-party computation
%enjoys a vast amount of publications in the
%classical computer science literature (as for instance
%\cite{Abadi2}\cite{Cachin}\cite{Yao} to cite only a few), this is
%not on the other hand the case for blind computation,
%which suffer as their most influential result a no-go
%theorem by Abadi et Al. \cite{Abadi1} together with a general
%disbelief that this can be achieved.
%There was a short regain of
%interest when Sander and Tschudin \cite{Sander} gave a blind
%computation protocol for evaluating polynomials relying on
%homomorphic encryption schemes, themselves based on computational
%assumptions.

%To our knowledge there has not been any radical
%development upon these matters since.\\
In this paper as in\cite{Feigenbaum,Abadi1}, we are 
concerned with unconditional security, 
that is we do not 
make any computational assumptions upon eavesdropper Bob.  
Although we give Bob the opportunity to gain some Shannon 
information $I$ about Alice's input $x$, we ensure that 
Bob's eavesdropping gets detected by Alice with a 
probability which rapidly increases with $I$. Any server 
Bob who wants to remain in business should clearly avoid 
such an a posteriori detection. Our goal consists of 
finding protocols for blind computation for which a good 
tradeoff between Bob's ability to avoid being detected and 
the amount of Shannon information about Alice's input can 
be established. Almost privacy was recently studied by 
Klauck \cite{Klauck} in a two-party computation setting 
which differs from the asymmetric scenario imposed by blind 
computations. Moreover, the security was only considered 
with respect to passive adversaries. We want our solution 
to apply to a wider class of functions than the one 
considered in the classical setting while being resistant 
to active adversaries.
%The uneven distribution of articles in quantum information is
%merely a refection of the situation in the non-quantum computer
%science literature. Since the birth of quantum cryptography
%\cite{BB84} much effort has gone, on the one hand, into devising
%secure multi-party quantum computation protocols. In particular
%scientists have long sought to provide unconditional quantum bit
%commitment and oblivious transfer, the primitives upon which
%multi-party computation is founded in the classical realm. The
%no-go theorem for quantum bit commitment by
%\cite{LoChau,Mayers} has shattered much of these hopes however, with the
%consequences explained in \cite{Lo}. Yet there has been a few
%results since, with more than two players \cite{Smith}, for some
%specific cryptographic tasks \cite{Klauck}, or using computational
%assumptions which are thought to resist quantum computational
%power \cite{Salvail}.
As far as we can tell, blind quantum computation has not 
been studied as such so far.

In section \ref{principles of solution} we present the 
basic ideas of our blind quantum computation protocol, as 
well as the reasons which limit their use to a certain 
class of functions. In section \ref{decoys review} we 
review and adapt a recent result in the Information versus 
Disturbance tradeoff literature. In section \ref{protocol 
and security} we formalize the protocol and give a proof of 
its security. We conclude in section \ref{conclusion bqc} 
and mention possible extensions of this work.

\section{Principles of a Solution} \label{principles of solution}
Let us now explain the basic principles underlying our 
blind quantum computation protocol. Suppose Alice wants Bob 
to compute $f(x)$ whilst keeping $x$ secret. Moreover 
suppose Bob possesses a quantum computer which implements 
$f$, i.e. $\!$he is able to implement a unitary transform 
$U$ such that $U\ket{q}\ket{0}\!=\!\ket{q}\ket{f(q)}$ for 
all inputs, $q$. In order to achieve her purpose Alice 
could hide her true input $\ket{x}$ amongst superpositions 
of other potential inputs 
$\frac{\ket{q}+i\ket{q'}}{\sqrt{2}}$ (which we later refer 
to as `quantum decoys') and send all this to Bob for him to 
execute $U$. Now if Bob attempts a measure so as to 
determine $\ket{x}$ he will run the risk of collapsing the 
superpositions. Alice may detect such a tampering when she 
retrieves her results. The above suggestion has a weakness 
however: Alice is not returned 
$\frac{\ket{q}+i\ket{q'}}{\sqrt{2}}$, but
\begin{equation*}
U\frac{\ket{q}+i\ket{q'}}{\sqrt{2}}\ket{0}=\frac{\ket{q;f(q)}+i\ket{q';f(q')}}{\sqrt{2}},
\end{equation*}
the result of Bob's computation upon the superposition 
Alice had sent. Since Alice does not want to compute $f$ 
herself she is in general unable to check upon the 
integrity of such states. To get an intuition of why this 
is consider the effects of tracing out the result register 
whenever $f(q)$ is different from $f(q')$.
\begin{align*} 
&\trace_2(\frac{\ket{q;f(q)}+i\ket{q';f(q')}\bra{q;f(q)}-i\bra{q';f(q')}}{2} 
)\\ &=\big(\ket{q}\bra{q}\trace(\ket{f(q)}\bra{f(q)}) 
-i\ket{q}\bra{q'}\trace(\ket{f(q)}\bra{f(q')})\\ 
&+i\ket{q'}\bra{q}\trace(\ket{f(q')}\bra{f(q)}) 
+\ket{q'}\bra{q'}\trace(\ket{f(q')}\bra{f(q')})\big)/2\\ 
&=\frac{\ket{q}\bra{q}+\ket{q'}\bra{q'}}{2}
\end{align*}
In other words once such a trace-out has been performed the 
state is either $\ket{q}$ with $0.5$ probability, or 
$\ket{q'}$ with $0.5$ probability, i.e. it makes no 
difference whether Bob performed a measurement in the 
computational basis or not.

\noindent There are many computational problems, however, 
for which this obstacle can be circumvened. For example say 
$f$ takes composite numbers into the list of their integer 
factors. Then Alice can easily (at the cost of a few 
multiplications) prepare several input-output pairs 
$\{(q,f(q))\}$. Thus if Alice hides her true input 
$\ket{x}$ amongst superpositions 
$\frac{\ket{q}+i\ket{q'}}{\sqrt{2}}$ generated in this 
manner, she will later be able to check whether 
$\frac{\ket{q;f(q)}+i\ket{q';f(q')}}{\sqrt{2}}$ are indeed 
being returned. Formally the idealized class of functions 
for which our protocol will work is defined as follows: 
\begin{Def}[Random verifiable functions] Let $S$ and $S'$ 
denote two finite sets. A function $f: S\rightarrow S'$ is 
{\em random verifiable} if and only if there exists, for 
all $N$, an efficient probabilistic process which generates 
$N$ input-output pairs $\{(q,f(q))\}$ and such that the 
inputs (the $q$'s) are uniformly distributed in $S$.
\end{Def}

There are several promised problems for which we can define 
functions that are random verifiable. Consider the language 
$RSA$-composite which contains natural numbers of a fixed 
size that can be expressed by the product of two primes of 
the same size. The function $f$ that returns the prime 
factors is also random verifiable. In this case, $f$ can be 
computed efficiently on a quantum computer but not, as far 
as we know, on a classical computer. Another example can be 
obtained from the graph isomorphism problem. Let $L_{e,v}$ 
be the set of all pairs of isomorphic graphs with $e$ edges 
and $v$ vertices. We define function $f:L_{e,v}\mapsto 
S_e$, where $S_e$ is the set of all permutations among $v$ 
elements, as $f(G_0,G_1)=\sigma$ such that 
$\sigma(G_0)=G_1$. It is easy to verify that $f$ is random 
verifiable. The following efficient classical computation 
does the job:
\begin{itemize}
\item Pick a random permutation $\sigma\in S_e$,
\item Generate a random graph $G_0$ with $e$ edges and $v$ 
vertices,
\item Output $((G_0,\sigma(G_0)),\sigma)$.
\end{itemize}
Although $f$ is random verifiable by an efficient classical 
algorithm, it is not known  whether even a quantum computer 
can evaluate $f$ efficiently.

%Even promised problems derived from NP-hard problems
%like Hamiltonian can be expressed that way. Suppose Bob's algorithm
%accepts a graph with at least one Hamiltonian cycle and returns it to Alice.
%If the number of edges and vertices are fixed, then Alice can
%run an algorithm very similar to the one for graph isomorphism
%in order to generate random graphs together with their Hamiltonian.

In this paper, we provide a blind quantum computation 
protocol for random verifiable functions together with a 
thorough security analysis. The cheat-sensitive security 
achieved relies upon the laws of physics only. It is 
expressed using the vocabulary of information theory. 
As 
was hinted in this section our analysis will crucially 
depend 
upon the tradeoff between Bob's information gain 
about Alice's true input (a canonical basis state) and the 
disturbance he induces upon superpositions of potential 
inputs (pairwise superpositions of canonical basis states).

\section{Information Gain versus Disturbance Tradeoff}\label{decoys review}
Say Alice draws out a state from an ensemble of quantum 
states, sends it to Bob, and later retrieves it. How much 
information can Bob learn about the state, and what, then, 
is the probability that Alice can detect Bob's 
eavesdropping? Questions of Information Gain versus 
Disturbance tradeoff were first investigated by Fuchs and 
Peres \cite{Fuchs}, who considered a seemingly simple 
scenario involving two equiprobable non-orthogonal pure 
states. But the formula they obtained is relatively complex 
and the methods employed are somewhat difficult to export 
to our setting. In order to construct a blind quantum 
computation protocol we needed to quantify the disturbance 
upon pairwise superpositions of $n$-dimensional canonical 
basis states, as induced when Bob seeks to learn 
information about the canonical basis. A tradeoff formula 
for this problem was given in \cite{decoys}. Proposition 
\ref{qdecoys} rephrases this result in terms of induced 
fidelity and letting Bob and Alice be the same person.

\begin{Scenario}[One quantum decoy]\label{qdecoys sc}
Consider a quantum channel for transmitting $n$-dimensional 
systems having canonical orthonormal basis $\{\ket{j}\}$.\\ 
Suppose Alice's message words are drawn out of the 
canonical ensemble $\{(1/n, \ket{j})\}_{j=1..n}$, whilst 
her \emph{quantum decoys} are drawn out of the pairing 
ensemble $\{(1/n^2,\frac{\ket{j}+i\ket{k}}{\sqrt{2}})\}$. 
Alice sends, over the quantum channel, either a message 
word or a decoy, which she later retrieves.\\ Whenever she 
sends a quantum decoy $\frac{\ket{j}+i\ket{k}}{\sqrt{2}}$ 
she later measures the retrieved system with $\{P_{intact}= 
\big( \frac{\ket{j}+i\ket{k}} {\sqrt{2}}\big) 
\big(\frac{\bra{j}-i\bra{k}}{\sqrt{2}}\big)\, ,\, 
P_{tamper}=\mathbb{I}-P_{intact}\}$ so as to check for 
tampering.\\ Suppose Bob is eavesdropping the quantum 
channel, and has an interest in determining Alice's message 
words.
\end{Scenario}

\begin{figure}[!h]
\center{\fbox{\begin{minipage}[t]{7.8cm}\vspace{-2mm}
\caption{\label{formal1}\small\textsc{One quantum decoy.}}\vspace{-3mm}
\begin{align*}
~\;\; A&:\; \textrm{Draw}\;t\;\textrm{in}\; 
T=\{(p,go),(1-p,nogo)\}\\ A&:\; 
\textrm{If}\;t=go\;\textrm{draw}\;s=m\;\textrm{in}\; 
M=\{(1/n, \ket{j})\}_{j=1..n}\\ A&:\; 
\textrm{If}\;t=nogo\;\textrm{draw}\;s=d\;\textrm{in}\; 
D=\{(1/n^2,\frac{\ket{j}+i\ket{k}}{\sqrt{2}})\}\\ A&:\; 
s\lra B\\ B&:\; \textrm{Draw}\;x,s'\;\textrm{in}\; 
X,S=\{(||M_x s||^2 , \ket{x}\otimes M_x s / ||M_x s||\}_x\\ 
&\textrm{with}\;\{M_x\}\;\textrm{a generalized 
measurement.}\\ B&:\; s'\lra A\\ A&:\; 
\textrm{If}\;t=go\;\textrm{draw}\;y\;\textrm{in}\; 
Y=\{(0,tamp),(1,notamp)\}\\ A&:\; 
\textrm{If}\;t=nogo\;\textrm{draw}\;y\;\textrm{in}\;\\ 
&Y=\{(||P_{tamper}s'||^2,tamp),(||P_{intact}s'||^2,notamp) 
\}\\ 
&\textrm{with}\;P_{tamper}=\mathbb{I}-P_{intact},\;P_{intact}=ss^{\dagger}.\\
%&\textrm{SECURITY PROPERTY:}\\
%&\textrm{Let}\;
%G=p(x=m|t=go)\;\textrm{and}\;F=p(y=notamp|t=nogo).\\
%&\textrm{Then}\;F\leq\frac{1}{2}+\frac{1}{2n}\Big(\sqrt{G}+\sqrt{(n-1)(1-G)}\Big)^2.
\end{align*}
\end{minipage}}}
\end{figure}

\begin{Proposition}[One quantum decoy]\label{qdecoys}~\\ 
Referring to Scenario \ref{qdecoys sc} and its 
formalization in Figure \ref{formal1}, suppose Bob performs 
an attack such that, whenever a message word gets sent, he 
is able to identify which with probability $G$ (mean 
estimation fidelity).\\ Then, whenever a quantum decoy gets 
sent, the probability $F$ (induced fidelity) of Bob's 
tampering not being detected by Alice is bounded above 
under the following tight inequality:
\begin{equation}\label{qdecoys eq}
F\leq\frac{1}{2}+\frac{1}{2n}\Big(\sqrt{G}+\sqrt{(n-1)(1-G)}\Big)^2
\end{equation}
For optimal attacks $G$ varies from $\frac{1}{n}$ to $1$ as 
$F$ varies from $1$ to $\frac{1}{2}\!+\!\frac{1}{2n}$.
\end{Proposition}

Now imagine that Scenario \ref{qdecoys sc} gets repeated 
$N$ times round, and that Alice happens to send only 
decoys. \begin{Scenario}[$N$ Quantum decoys]\label{nqdecoys 
sc}~\\ Step 0. Alice prepares a pool of $N+1$ quantum 
states consisting of one message word together with $N$ 
quantum decoys.\\ Step 1. Alice sends Bob one quantum state 
drawn at random amongst those remaining in the pool.\\ Step 
2. Alice awaits to retrieve the quantum state she sent.\\ 
Step 3. If Alice sent a quantum decoy she measures the 
retrieved system so as to check for tampering.\\
%Whenever she detects such a tampering she stops.\\
Step 4. If the pool is empty Alice stops the protocol, else 
she proceeds again with Step 1.\\ Step 5. Alice publicly 
announces the position $p$ at which she sent her message 
word.\vspace{0.1cm}\\ This scenario is formalized in Figure 
\ref{formal indiv}.\\
\end{Scenario}

\begin{figure}[!h]
\center{\fbox{\begin{minipage}[t]{8cm}\vspace{-2mm}
\caption{\label{formal indiv}\small\textsc{$N$ quantum 
decoys / individual attacks}}\vspace{-3mm}
\begin{align*}
~\; A:\; &\textrm{draw}\;p\;\textrm{in}\; 
P=\{(1/(N+1),i),\}_{i=0\ldots N}\\ \textrm{For}&\;r=0\ldots 
N:\\ &A:\; \textrm{If}\;r\neq 
p\;\textrm{draw}\;t\;\textrm{in}\; T=\{(1,go),(0,nogo)\}\\ 
&\quad \textrm{else draw}\;t\;\textrm{in}\; 
T=\{(0,go),(1,nogo)\}\\ &A:\; \textrm{If 
}\;t=go\;\textrm{draw }\;s=m\;\textrm{in}\; M=\{(1/n, 
\ket{j})\}_{j=1..n}\\ &A:\; \textrm{If 
}\;t=nogo\;\textrm{draw }\;s=d\;\textrm{ in}\\ &\quad 
D=\{(1/n^2,\frac{\ket{j}+i\ket{k}}{\sqrt{2}})\}\\ &A:\; 
s\lra B\\ &B:\; \textrm{Draw }\;x_r,s'\;\textrm{ in}\;\\ 
&\quad X,S=\{(||M^{(r)}_x s||^2 , \ket{x}\otimes M^{(r)}_x 
s/ ||M^{(r)}_x s||\}_r\\ &\quad\textrm{with, for all 
}\;r,\;\{M^{(r)}_x\}\;\\ &\quad\textrm{a gen. mesurement 
upon a bipartite system.}\\ &B:\; s'\lra A\\ &A:\; 
\textrm{If }\;t=go\;\textrm{draw }\;y_r\;\textrm{in}\\ 
&\quad Y=\{(0,tamp),(1,notamp)\}\\ &A:\; \textrm{If 
}\;t=nogo\;\textrm{ draw }\;y_r\;\textrm{ in}\;\\ &\quad 
Y=\{(||P_{tamper}s'||^2,tamp),(||P_{intact}v||^2,notamp) 
\}\\ &\quad \textrm{with 
}\;P_{tamper}=\mathbb{I}-P_{intact}, 
P_{intact}=ss^{\dagger}.\\ A:\; &p\lra B\\ B:\; 
&guess=f_p(x_0,\ldots,x_N)\\
\end{align*}
\end{minipage}}}
\end{figure}

\begin{Cor}[$N$ quantum decoys/individual 
attacks]\label{nqdecoys} Referring to Referring to Scenario 
\ref{nqdecoys sc}, suppose Bob only performs individual 
attacks, i.e. independent of each other at each round, as 
formalized in Figure \ref{formal indiv}.\\ The probability 
of Bob reaching round $m$ ($0\leq m \leq N$) without being 
caught tampering is bounded above under the following tight 
inequality:\\
\begin{equation}
p(\textrm{Bob reaches $m$})\leq\prod_{i=1,i\neq 
p}^{m-1}F(G_i)\label{interactive fidelity eq}
\end{equation}
where $G_i$ stands for Bob's mean estimation fidelity, if 
$p$ is announced equal to $i$, about the message word sent 
at round $p$.
\end{Cor}
\textbf{Proof.} Within the for loop of Figure \ref{formal 
indiv} the scenario which gets repeated is exactly that of 
Figure \ref{formal1}, for which Proposition \ref{qdecoys} 
applies independently at each round.\hfill $\Box$

In the above scenario Bob's attacks are somewhat 
memoryless. Bob's measurements do not depend upon previous 
outcomes nor upon any ancilla quantum system which he might 
keep throughout the protocol. This is what enables us to 
apply Proposition \ref{qdecoys} at the level of each 
individual transmission \cite{decoys}, i.e. to assume that 
that the probabilities $\{p(\textrm{Bob passing round 
}i)=F(G_i)\}_{i=1..N}$ are independent from each other and 
hence that that Bob's chances of not being detected at all 
are bounded by $\prod_{i=1..N} F(G_i)$.

\noindent Now say Bob was to keep an ancillary quantum 
system entangled with a quantum decoy sent at a previous 
round, and then perform a coherent quantum measurement upon 
another quantum decoy and the ancillary quantum system at a 
later round -- could this correlate his probabilities of 
getting caught in a favorable manner? We argue that it is 
not so in the following conjecture, by making use of a 
standard argument. Formal proofs of probabilistic security 
protocols are known to be an extremely delicate matter 
requiring delicate notions of process equivalences. In 
quantum information theory such rigorous frameworks have 
not yet appeared and seem to be needed here -- we will only 
provide the reader with a number of intuitions which 
strongly support our statement.

%However it is clear that if
%Scenario \ref{qdecoys sc} is repeated \emph{conditionally} upon
%Alice's measurement outcome, then we can no longer have such
%correlations:

\begin{figure}[!h]
\center{\fbox{\begin{minipage}[t]{8cm}\vspace{-2mm}
\caption{\label{formal coh}\small\textsc{$N$ quantum decoys 
/ coherent attacks}}\vspace{-3mm}
\begin{align*}
~\; A:\; &\textrm{draw}\;p\;\textrm{in}\; 
P=\{(1/(N+1),i),\}_{i=0\ldots N}\\ B:\; &a=\psi\\ 
\textrm{For}&\;r=0\ldots N:\\ &A:\; \textrm{If}\;r\neq 
p\;\textrm{draw}\;t\;\textrm{in}\; T=\{(1,go),(0,nogo)\}\\ 
&\quad \textrm{else draw}\;t\;\textrm{in}\; 
T=\{(0,go),(1,nogo)\}\\ &A:\; \textrm{If 
}\;t=go\;\textrm{draw }\;s=m\;\textrm{in}\; M=\{(1/n, 
\ket{j})\}_{j=1..n}\\ &A:\; \textrm{If 
}\;t=nogo\;\textrm{draw }\;s=d\;\textrm{ in}\\ &\quad 
D=\{(1/n^2,\frac{\ket{j}+i\ket{k}}{\sqrt{2}})\}\\ &A:\; 
s\lra B\\ &B:\; \textrm{Draw }\;x_r,s',a'\;\textrm{ in}\;\\ 
&\quad X,S,A=\{(||M^{(r)}_x s a||^2 , \ket{x}\otimes 
M^{(r)}_x s a / ||M^{(r)}_x s a||\}_r\\ &\quad\textrm{with, 
for all }\;r,\;\{M^{(r)}_x\}\;\\ &\quad\textrm{a gen. 
mesurement upon a bipartite system.}\\ &B:\; s'\lra A\\ 
&A:\; \textrm{If }\;t=go\;\textrm{draw 
}\;y_r\;\textrm{in}\\ &\quad Y=\{(0,tamp),(1,notamp)\}\\ 
&A:\; \textrm{If }\;t=nogo\;\textrm{ draw }\;y_r\;\textrm{ 
in}\;\\ &\quad 
Y=\{(||P_{tamper}s'||^2,tamp),(||P_{intact}v||^2,notamp) 
\}\\ &\quad \textrm{with 
}\;P_{tamper}=\mathbb{I}-P_{intact}, 
P_{intact}=ss^{\dagger}.\\ &B:\; a=a'\\ A:\; &p\lra B\\ 
B:\; &guess=f_p(x_0,\ldots,x_N)\\
\end{align*}
\end{minipage}}}
\end{figure}

\begin{Conj}[$N$ quantum decoys/coherent attacks]\label{conj}
Referring to Scenario \ref{nqdecoys sc}, suppose Bob 
performs general attacks, i.e. which may depend from each 
other at every round, as formalized in Figure \ref{formal 
coh}.\\ The probability of Bob reaching round $m$ ($0\leq m 
\leq N$) without being caught tampering is bounded above 
under the following tight inequality:\\
\begin{equation*}
p(\textrm{Bob reaches $m$})\leq\prod_{i=1,i\neq 
p}^{m-1}F(G_i)
\end{equation*}
where $G_i$ stands for Bob's mean estimation fidelity, if 
$p$ is announced equal to $i$, about the message word sent 
at round $p$.
\end{Conj}
\emph{The following arguments support our claim.}   
% ONE LAST ATTEMPT: SHOW a IS UNCORRELATED AND THEREFORE
% NON-INFLUENT ACCORDING TO SOME INFO THEORETICAL ARGUMENT.
% MEANWHILE...
Note that in this Figure \ref{formal coh} we allow Bob to 
perform the most general attack possible: his generalized 
measurements $\{M^{(r)}_x\}$ depend upon the round $r$; 
they may entangle the ancillary quantum system $a$ to the 
state sent by Alice $s$ for later use (thus the systems 
$a'$ and $s'$ may be entangled); they may depend upon 
previous measurement outcomes via the contents of the 
ancillary quantum system $a$; or they could keep $a$ 
entangled but unmeasured until the final round provided 
that for $r<N$ the statistics of $\{M^{(r)}_x\}$ do not 
depend on $a$. We now reason by contradiction.\\ Suppose 
$p(\textrm{Bob reaches }m)>\prod_{i=1,i\neq 
p}^{m-1}F(G_i)$. Then there exists a $k$ for which
\begin{align*}
p(\textrm{Bob reaches }k)&\leq\prod_{i=1,i\neq 
p}^{k-1}F(G_i)\quad\textrm{and}\\ p(\textrm{Bob reaches 
}k\!+\!1)&>\prod_{i=1,i\neq p}^{k}F(G_i).
\end{align*}
For such a $k$ we thus have
\begin{equation}\label{interactive fidelity contradiction}
p(\textrm{Bob reaches }k\!+\!1|\textrm{Bob reaches 
}k)>F(G_k).
\end{equation}
In other words Bob, on the $k^{th}$ round, due to the state 
of the ancillary system $a$ at this round, is capable of 
collecting mean estimation $G_k$ about a message word 
whilst remaining undetected with probability more than 
$F(G_k)$ upon a quantum decoy. However $a$ is absolutely 
uncorrelated with $s$ for our purpose, because:
\begin{itemize}
\item the quantum decoys and the message words are 
undistinguishable since $(1/n^2)\sum_{jk} 
(\ket{j}+i\ket{k})(\bra{j}-i\bra{k})=(1/n)\sum_i 
\ket{i}\bra{i}$. Hence $a$ cannot hold any information 
about whether the message word is sent at round $k$;
\item the quantum decoys are picked up independently from 
one another and independently from the message words, hence 
if the message word is sent at round $k$, $a$ does not hold 
any complementary information about the message word and 
does not modify $G_k$;
\item the quantum decoys are picked up independently from 
one another and independently from the message words, hence 
if a quantum decoy is sent at round $k$, $a$ does not hold 
any complementary information about the subspace of the 
quantum decoy which needs to be preserved and hence does 
not modify $F(G_k)$.
\end{itemize}
In other words Bob could have, for the purpose of 
optimizing his information gain versus disturbance tradeoff 
at round $k$, come up with just as good an $a$ by playing 
the first $k-1$ rounds of the protocol with Charlie 
instead. Hence the situation at round $k$ is in 
contradiction with Proposition 
\ref{qdecoys}.\hfill$\boxtimes$\\

The next section also makes use of the following 
mathematical result, whose direct proof was shown to us by 
Prof. Frank Kelly.
\begin{Lem}[Concavity of circular products]\label{Kelly}
Consider $f:[0,1]\rightarrow[0,1]$ a concave, continuous 
function and $\{x_i\}_{i=1\ldots N+1}$ a set of real 
numbers in the interval $[0,1]$.\\ Suppose the sum 
$t=\sum_{i=1}^{N+1} x_i$ is fixed. We have
\begin{equation*}
\frac{1}{N+1}\sum_{p=1}^{N+1}\Big(\prod_{i=1,i\neq 
p}^{i=N+1}f(x_i)\Big)\leq{f\big(\frac{t}{N+1}\big)}^{N}.
\end{equation*}
\end{Lem}
\emph{Proof.} By definition of concavity one has
\begin{align}
\frac{1}{2}\big(f(x_1)+f(x_2)\big)&\leq 
f\big(\frac{x_1+x_2}{2}\big)\label{sum concavity eq}\\ 
\textrm{and}\quad f(x_1)f(x_2)&\leq 
{f\big(\frac{x_1+x_2}{2}\big)}^2\label{product concavity 
eq},
\end{align}
where the latter equation trivially derives from 
$f(x_1)f(x_2)\leq\big(\frac{f(x_1)+f(x_2)}{2}\big)^2$. Let 
us now show that
\begin{equation}\label{pairwise average eq}
\frac{1}{N+1}\sum_{p=1}^{N+1}\prod_{i=1,i\neq 
p}^{i=N+1}f(x_i)\leq\frac{1}{N+1}\sum_{p=1}^{N+1}\prod_{i=1,i\neq 
p}^{i=N+1}f(y_i),
\end{equation}
where $y_1=y_2=\frac{x_1+x_2}{2}$ and $y_i=x_i$ for 
$i=3\ldots N+1$. This result is in fact obtained by 
combining (summing) two inequalities:
\begin{align*}
\big(f(x_1)+f(x_2)\big)\!\prod_{i=3}^{N+1}f(x_i)&\leq 
\big(f(y_1)+f(y_2)\big)\!\prod_{i=3}^{N+1}f(y_i)\\ 
f(x_1)f(x_2)\!\sum_{p=3}^{N+1}\prod_{i=3,i\neq 
p}^{i=N+1}\!\! f(x_i)&\leq 
f(y_1)f(y_2)\!\sum_{p=3}^{N+1}\prod_{i=3,i\neq 
p}^{i=N+1}\!\! f(y_i)
\end{align*}
where former stems from Equation (\ref{sum concavity eq}) 
and $f(x)$ positive, whilst the latter stems from Equation 
(\ref{product concavity eq}) and $f(x)$ positive.

Equation (\ref{pairwise average eq}) expresses the fact 
that, whenever two elements $x_i$ and $x_j$, $i\neq j$ are 
replaced by their mean, the value of
\begin{equation*}
\pi(\ul{x})\equiv\frac{1}{N+1}\sum_{p=1}^{N+1}\Big(\prod_{i=1,i\neq 
p}^{i=N+1}f(x_i)\Big)
\end{equation*}
is increased. Now let us define $\{\ul{x}^{(k)}\}$ a 
sequence of vectors such that $\ul{x}^{(1)}=(x_1, 
x_2,\ldots ,x_{N+1})$, and $\ul{x}^{(k)}$ is formed from 
$\ul{x}^{(k-1)}$ by replacing both the largest and the 
smallest component by their mean. As $k$ goes to infinity 
this sequence of vectors tends to 
$\ul{x}^{(\infty)}=(\frac{t}{N+1}, \frac{t}{N+1},\ldots)$. 
By Equation (\ref{pairwise average eq}) we have 
$\{\pi(\ul{x}^{(k)})\}$ an increasing sequence of real 
numbers. As $k$ goes to infinity, and since $\pi(\ul{x})$ 
is continuous in $\ul{x}$, this sequence of real numbers 
tends to
\begin{equation*}
\pi(\ul{x}^{(\infty)})=f\big(\frac{t}{N+1}\big)^{N}.
\end{equation*}
This limit must therefore provide, for all $\ul{x}$ having 
components summing to $t$, a tight upper bound on the value 
of $\pi(\ul{x})$.\hfill$\quad\Box$\\

\section{Protocol and Security}\label{protocol and security}
We are now set to give our blind quantum computation 
protocol:
\begin{Protocol}[Interactive version]\label{interactive protocol}
Alice wants Bob to compute $f(x)$ whilst keeping her input 
$x$ secret. Here $f$ designates a random verifiable 
function implemented on a quantum computer by a unitary 
evolution $U$.\vspace{0.1cm}\\ Step 0. Alice efficiently 
computes $2N$ random input-solution pairs $(q,f(q))$ and 
prepares a pool of $N+1$ quantum states consisting of her 
true input $\ket{x}$ together with $N$ quantum decoys 
$\frac{\ket{q}+i\ket{q'}}{\sqrt{2}}$.\\ Step 1. Alice sends 
Bob one quantum state $\ket{\psi}$ drawn at random amongst 
those remaining in the pool.\\ Step 2. Bob supposedly 
computes $U\ket{\psi}\ket{0}$ and sends the result back to 
Alice.\\ Step 3. If $\ket{\psi}$ was a quantum decoy 
$\frac{\ket{q}+i\ket{q'}}{\sqrt{2}}$ Alice measures the 
retrieved system with
\begin{align*}
\big\{P_{intact}&= \frac{1}{2}\big( \ket{q\,f(q)}+i\ket{q' 
f(q')}\big) \big(\bra{q\,f(q)}-i\bra{q' f(q')}\big)\\ 
P_{tamper}&=\mathbb{I}-P_{intact}\big\},
\end{align*}
so as to check for tampering. % Whenever she detects such a 
tampering she stops. 
If on the other hand $\ket{\psi}$ was 
her true input Alice reads off $f(x)$.\\ Step 4. If the 
pool is empty Alice stops the protocol, else she proceeds 
again with Step 1. \end{Protocol}
Quantum theory is helpful 
for detecting observation by a mistrusted party through the 
induced disturbance. For this reason quantum cryptography 
has seen the rise of \emph{cheat-sensitive} protocols where 
`Either party may be able to evade the intended constraints 
on information transfer by deviating from these protocols. 
However, if they do, there is a non-zero probability that 
the other will detect their cheating' \cite{Hardy}.  When 
the probability of detecting the cheating is one, the 
protocol may also be referred to as \emph{cheat-evident} 
\cite{Colbeck}.\\ The security of our protocol is 
cheat-sensitive, as is rigorously described and quantified 
in the following claim.\\
The security of our protocol may 
also be referred to as cheat-evident, in the sense that 
Alice's detection probability tends to $1$ in the limit 
where $N$ tends to infinity. Moreover for a fixed 
information gain by Bob, Alice's detection probability 
approaches $1$ exponentially with $N$. 

\begin{Cl}[Statement of security]\label{claim security bqc}
Referring to Protocol \ref{interactive protocol} suppose 
Bob has no a priori information about Alice's true input 
$x$. Let $I\in[0,\log(n)]$ be Bob's mutual information 
about Alice's true input $x$ at the end of the protocol. 
Let  $D\in[0,(1/2)^N]$ be the probability of Alice 
detecting Bob's tampering. Provided that Bob makes only 
individual attacks, the protocol ensures that $\forall 
G\in[\frac{1}{n},1]$, 
\begin{align*}
\big[I=\log(n)+\log(G)\;\;\Rightarrow\;\; 
D\geq 1-F(G)^{N}\big].
\end{align*}
Hence we have equivalently
\begin{align*} D\geq 
1-F(2^{I-\log(n)})^{N}. \end{align*}
\end{Cl} \emph{Proof.} 

%defining G_i and \ov{G} ---------------
We prove that the 
claim holds for a weakened form of Protocol 
\ref{interactive protocol}, where we add:\\ \emph{Step 5. 
Alice publicly announces the position in which she sent her 
true input $\ket{x}$.}\\ Until this stage, however, Bob has 
no means of knowing at which round true input $\ket{x}$ was 
sent. This is because we have assumed he has no a priori 
knowledge about the true input. In his view the state was 
drawn from the canonical ensemble $\{(1/n, 
\ket{j})\}_{j=1..n}$, whilst the quantum decoys were drawn 
from the pairing ensemble 
$\{(1/n^2,\frac{\ket{j}+i\ket{k}}{\sqrt{2}})\}$, but the 
two are undistinguishable for they both have density matrix 
$\mathbb{I}/n$. We are, therefore, in the precise case of 
Corollary \ref{nqdecoys}. Without loss of generality we can 
assume Bob's attack yields him mean estimation fidelity 
$G_i$ about Alice's true input whenever the position is 
later announced equal to $i$. Let $\ov{G}=\sum_p G_p/(N+1)$.

%bound on information ---------------
\noindent First we prove that 
$[I=\log(n)+\log(G)\Rightarrow \ov{G}\geq G ]$.\\
Say the 
true input is at position $p$. In this situation Bob's best 
chance of guessing the true input is $G_p$ (by definition) 
and thus his Shannon uncertainty $H_p$ about Alice's true 
input is bounded as follows
\begin{align*}
H_p&\equiv \sum -p(x|\textrm{Bob's 
outcome})\log(p(x|\textrm{Bob's outcome}))\\ &\geq-\lfloor 
\frac{1}{G_p} \rfloor G_p 
\log(G_p)-(1-\lfloor\frac{1}{G_p}\rfloor 
G_p)\log(1-\lfloor\frac{1}{G_p}\rfloor G_p)\\ &\geq 
-\log(G_p).
\end{align*}
The RHS of the last line is often referred to as the 
`min-entropy' sometimes denoted $H_{\infty}$ and is 
commonly used to bound uncertainties in the above manner 
(i.e. Shannon uncertainty is always at least $H_{\infty}$). 
As a consequence Bob's mutual information $I_p$ satisfies
\begin{align*}
I_p\leq \log(n)+\log(G_p).
\end{align*}
Averaging over all possible positions $p=1\ldots N+1$ Bob's 
mutual information satisfies
\begin{align*}
I&=\sum_{p=1}^{N+1}\frac{1}{N+1}I_p\\ &\leq 
\log(n)+\sum_{p=1}^{N+1}\frac{1}{N+1}\log(G_p)\\ &\leq 
\log(n)+\log(\ov{G})
\end{align*}
where the third line was obtained using the concavity of 
$x\mapsto\log(x)$. Hence we have
\begin{align*} 
I=\log(n)+\log(G)\leq \log(n)+\log(\ov{G}). 
\end{align*}
Since $x\mapsto\log(x)$ is crescent we 
conclude that $G\leq\ov{G}$.
%bound on disturbance ---------------------
\noindent Second we prove that $[\ov{G}\geq G\Rightarrow 
D\geq  1-F\big(G\big)^{N} ]$.\\
Since we have assumed 
individual attacks Corollary \ref{nqdecoys} applies, and so 
Bob is undetected with probability
\begin{align*}
p(\textrm{undetected}|p)&\leq\prod_{i=1,i\neq p}^{N+1} 
F(G_i).
\end{align*}
Let us now average the above over all possible positions 
$p=1\ldots N+1$. The probability that Bob's tampering 
remains undetected by Alice satisfies
\begin{align*}
p(\textrm{undetected})&=\frac{1}{N+1}\sum_{p=1}^{N+1}p(\textrm{undetected}|p)\\ 
&\leq\frac{1}{N+1}\sum_{p=1}^{N+1}\big(\prod_{i=1,i\neq 
p}^{N+1} F(G_i)\big)\\ &\leq F\big(\ov{G}\big)^{N}\\
D&\geq 
1-F\big(\ov{G}\big)^{N}.
\end{align*}
where the third line was obtained using Lemma \ref{Kelly} 
upon the concave, continuous function $x\mapsto 
F(x)$.
Since $x\mapsto  1-F\big(x\big)^{N}$ is crescent and 
$\ov{G}\geq G$ we conclude that
\begin{align*} D\geq 
1-F\big(\ov{G}\big)^{N}\geq 1-F\big(G\big)^{N}.
\end{align*}
%end the proof----------------------------
\hfill$\quad\Box$\\

Protocol \ref{interactive protocol} requires $N+1$ 
communications between Alice and Bob. One could suggest a 
modification whereby Alice would send Bob her whole pool 
(as prepared in \emph{Step 0}), and later proceed to check 
upon the integrity of each element of the pool which Bob 
returns, apart from her true input. Formally this yields 
the following protocol:
\begin{Protocol}[Non-interactive version]\label{noninteractive protocol}
Alice wants Bob to compute $f(x)$ whilst keeping her input 
$x$ secret. Here $f$ designates a random verifiable 
function implemented on a quantum computer by a unitary 
evolution $U$.\vspace{0.1cm}\\ Step 0. Alice efficiently 
computes $2N$ random input-solution pairs $(q,f(q))$ and 
prepares a pool of $N+1$ quantum states consisting of her 
true input $\ket{x}$ together with $N$ quantum decoys 
$\frac{\ket{q}+i\ket{q'}}{\sqrt{2}}$.\\ Step 1. Alice sends 
Bob the large quantum state 
$\bigotimes_{i=1}^{N+1}\ket{\psi_i}$ consisting of a random 
permutation of all elements of the pool.\\ Step 2. Bob 
supposedly computes $\bigotimes_{i=1}^{N+1} 
U\ket{\psi_i}\ket{0}$ and sends the result back to Alice.\\ 
Step 3. For each location $i$, if $\ket{\psi_i}$ was a 
quantum decoy $\frac{\ket{q}+i\ket{q'}}{\sqrt{2}}$ Alice 
measures
\begin{align*}
\big\{P_{intact}&= \frac{1}{2}\big( \ket{q;f(q)}+i\ket{q'; 
f(q')}\big) \big(\bra{q;f(q)}-i\bra{q'; f(q')}\big)\\ 
P_{tamper}&=\mathbb{I}-P_{intact}\big\}. \end{align*} so as 
to check for tampering. If on the other hand $\ket{\psi_i}$ 
was her true input Alice reads off $f(x)$.\\
\end{Protocol} 
When Bob is restricted to individual attacks (non-coherent 
attacks, i.e. Bob measures each quantum state in the pool 
individually) then Claim \ref{claim security bqc} holds 
also for Protocol \ref{noninteractive protocol}. We omit 
the proof of this since it is similar, and in fact simpler 
than the one given for Protocol \ref{interactive protocol}. 
Now suppose Conjecture \ref{conj} was verified. This would 
immediately entail that Claim \ref{claim security bqc} 
holds also for coherent attacks for Protocol 
\ref{interactive protocol}. Hence we believe that Protocol 
\ref{interactive protocol} can resist the most general 
attack. A thought-provoking question is whether this is 
still the case of Protocol \ref{noninteractive protocol}. 
Is it the case that interactivity contributes, to some 
extent, to a limitation of Bob's possible attacks?
 
\section{Concluding Remarks}\label{conclusion bqc}
We have investigated the possibility of \emph{having 
someone else carrying out the evaluation of a function for 
you without letting him learn anything about your input}. 
We gave a blind computation protocol for the class of 
functions which admit an efficient procedure to generate 
random input-output pairs. The protocol relies upon quantum 
physical information gain versus disturbance tradeoffs 
\cite{decoys} to achieve cheat-sensitive security against 
individual attacks: whenever the server gathers 
$\log(n)+\log(G)$ bits of Shannon information about the 
input, he must get caught with probability at least 
$1-F(G)^{N}$ (where $n$ denotes the size of the input and 
$N$ is a security parameter). Moreover the server cannot 
distinguish a weary client who uses the blind computation 
protocol (sending one true input amongst $N$ decoys) from a 
normal client who simply makes repeated use of the server 
(sending $N+1$ true inputs). Thus if the server wanted to 
deny his services to suspected users of the protocol, he 
would also have to refuse the normal clients. We have 
conjectured that the same security properties hold for 
general, coherent attacks.

Our protocol could be improved in several directions.\\ In 
terms of costs one may hope to reduce the set of quantum 
gates needed by Alice to prepare her transmissions 
\cite{Childs}; lower the size of the transmissions; lower 
the number of rounds required. We leave it as an open 
problem to find the security properties of the 
non-interactive version of our protocol when Bob is allowed 
coherent attacks.\\ In terms of functionality one may wish 
to achieve tamper prevention (preventing Bob from learning 
about $x$ ever) rather than tamper detection (preventing 
Bob from learning about $x$ without being detected, i.e. 
cheat-sensitiveness). Protocol \ref{interactive protocol} 
provides the latter to some degree, since its interactivity 
allows Alice to avoid sending her true input $x$ whenever 
she detects tampering upon her quantum decoys in the 
previous rounds. However we have not provided an analysis 
for a tamper-prevention-like security property. Another 
challenge would be to extend/identify the class of 
functions admitting a blind quantum computation protocol. 
This may have consequences in quantum complexity theory, as 
was the case in the classical setting \cite{Abadi1}. For 
instance if one was to prove that the blind quantum 
computation protocol had no interest as a secure way of 
discharging Alice computationally - because all the random 
verifiable functions turn out to be easy to perform on a 
quantum computer - then random verifiability would impose 
itself as an elegant property for the quantum polynomial 
class.

\section*{Acknowledgments} P.J.A  would like to thank Prof. 
Frank Kelly for his self-contained proof of Lemma 
\ref{Kelly}, Dr. Anuj Dawar for proof-reading, EPSRC, 
Marconi, the Cambridge European and Isaac Newton Trusts, 
and the European Union Marie Curie Fellowship scheme for 
financial support. L. S. would like to thank Ivan Damg\aa 
rd for enlightening discussions and the EU project PROSECCO 
for financial support. Both would like to thank the 
anonymous referee for important comments.

\end{document}